\documentclass[prl,twocolumn,floatfix,showpacs,superscriptaddress]{revtex4}

\usepackage[latin1]{inputenc}
\usepackage{amsmath}
\usepackage{mathtools}
\usepackage{mathrsfs }

\usepackage[usenames,dvipsnames]{color}
\usepackage{amsfonts}

\usepackage{amssymb,latexsym} 

\usepackage{enumerate}
\usepackage{soul}
\usepackage{graphicx}

\usepackage{hyperref}
\usepackage{multirow}
\usepackage{color}

\newcommand{\beq}[1]{\begin{equation}\label{#1}}
\newcommand{\eeq}{\end{equation}}

\begin{document}
\title{
Large-scale analysis of 
Zipf's law in English texts
}
\author{
Isabel Moreno-S\'anchez, Francesc Font-Clos, \'Alvaro Corral}
\affiliation{Centre de Recerca Matem\`atica,
Edifici C, Campus Bellaterra,
E-08193 Barcelona, Spain.
} 
\affiliation{Departament de Matem\`atiques,
Facultat de Ci\`encies,
Universitat Aut\`onoma de Barcelona,
E-08193 Barcelona, Spain}

\begin{abstract}
Despite being a paradigm of quantitative linguistics,
Zipf's law for words
suffers from three main problems:
its formulation is ambiguous, 
its validity has not been tested rigorously from a statistical point of view, 
and it has not been confronted to a representatively large number of texts.
So, we can summarize the current support of Zipf's law in texts
as anecdotic.
We try to solve these issues by studying three different versions of Zipf's law and
fitting them to all available English texts in the Project Gutenberg database
(consisting of more than 30\,000 texts). To do so we use state-of-the art tools in fitting and goodness-of-fit tests,
carefully tailored to the peculiarities of text statistics.
Remarkably, one of the three versions of Zipf's law, 
consisting of a pure power-law form in the complementary cumulative distribution
function of word frequencies, is able to fit more than 40\,\%
of the texts in the database (at the 0.05 significance level), 
for the whole domain of frequencies (from 1 to the maximum value)
and with only one free parameter (the exponent).
\end{abstract}

\date{\today}

\maketitle

\section{Introduction}

Zipf's law constitutes a striking quantitative regularity 
in the usage of language \cite{Baayen,Baroni2009,Zanette_book,Piantadosi}.
It states that, for a large enough piece of text, 
the frequency of use $n$ of any word decreases with its
rareness $r$ in the text in an approximately hyperbolic way, i.e., $n \propto 1/r$,
where the symbol ``$\propto$'' denotes proportionality. 
Technically, $r$ is called the rank, and the most common (i.e., less rare) word is assigned $r=1$, 
the second most common, $r=2$, and so on.
A slightly more general formulation includes a parameter in the form of an exponent $\alpha$; 
then, the rank-frequency relation takes the form of a power law,
\begin{equation}
n \propto \frac 1 {r^\alpha}.
\label{laprimera}
\end{equation}
with the value of $\alpha$ close to one.

This pattern (\ref{laprimera})
has been found across different languages, literary styles, time periods, and %at several 
levels of morphological abstraction \cite{Baroni2009,Zanette_2005,Corral_Boleda}.
More fascinatingly, the same law has been claimed in other codes of communication, as in 
music \cite{Serra_scirep}
or for the timbres of sounds \cite{Haro},
and also in disparate discrete systems where individual units or agents
gather into different classes \cite{Li02}, for example, 
employees into firms \cite{Axtell},
believers into religions \cite{Clauset},
insects into plants \cite{Pueyo},
units of mass into animals present in ecosystems \cite{Camacho_sole},
visitors or links into web pages \cite{Adamic_Huberman},
telephone calls to users \cite{Newman_05},
or abundance of proteins (in a single cell) \cite{Furusawa2003}.
The attempts to find an explanation have been diverse \cite{Simon,Miller_monkey,Ferrer2002a,Mitz,Newman_05,Saichev_Sornette_Zipf,Zanette_book,Corominas-Murtra_2011,Peterson_Dill,Corominas_dice},
but no solution has raised consensus \cite{Mitz,Cancho_monkey,Dickman_Moloney_Altmann}.

Despite its quantitative character, 
Zipf's law has been usually checked for in a qualitative way, 
plotting the logarithm of the frequency $n$
versus the logarithm of the rank $r$
and looking for some domain with a roughly linear behavior, 
with slope more or less close to $-1$. 
A more refined approach consists in fitting a straight line
to the double-logarithmic plot by linear regression
\cite{Li2010a}.
But several authors have recently pointed out the limitations
of this method when applied to probability distributions
\cite{Goldstein,Bauke,White,Clauset}, 
and the advantages of using an asymptotically unbiased and minimum-variance procedure 
such as maximum likelihood (ML) estimation \cite{White}, 
whose solutions, moreover, are invariant under reparameterizations \cite{Casella,Corral_Deluca}.
One should consider then ML estimation as the most reliable procedure
of estimation for parametric models 
(when a maximum of the likelihood does exist 
and the number of data is large).

Furthermore, for the particular case of linguistics, 
the search for Zipf's law has been traditionally performed in very limited sets
of texts (less than a dozen in a typical research article \cite{FontClos_Corral,Corral_Boleda}). 
More recently, however, large corpora have been considered
--these are representative collections of different texts aggregated together
into a single bag, so, instead of many separated texts one deals with one enormous mixed text.
When ``rare'' words are not considered
(i.e., for high frequencies), it seems that 
Zipf's law still holds in these large collections
\cite{Baroni2009,Ferrer2000a,Petersen_scirep,Gerlach_Altmann,Williams}.
At present, there is agreement that Zipf's law is a rough approximation
in lexical statistics, but its range of validity is totally unknown, 
i.e., we ignore how good Zipf's law is in order to account for the appearance of words, 
and for which texts it should work --and with which level of precision-- 
and for which texts it should fail.

An extra difficulty emerges when one recognizes the ill-defined nature of Zipf's law.
In fact, the law has two formulations, with the first one being Eq. (\ref{laprimera}), 
which just counts the frequency of words.
For the sake of clarity, the words that are counted are referred to as word types,
in order to distinguish them from  each repetition, 
which is called a token.
The second formulation of Zipf's law arises when, after counting the frequency of word types, 
one performs a second statistics and counts 
how many values of the frequency are repeated,
that is, 
how many word types have the same frequency.
This means that the frequency $n$ is considered the random variable.
One can realize that the rank, when normalized by its maximum value in text, 
is just the empirical estimation of the complementary cumulative distribution 
function of $n$, 
and then, the derivative of the expression for $r(n)$ (the inverse of Eq. (\ref{laprimera}))
yields a continuous approximation
for the probability mass function $f(n)$ of the frequency $n$.
From here one obtains another power law,
\begin{equation}
f(n) \propto \frac 1 {n^\beta},
\label{lasegunda}
\end{equation}
with the new exponent $\beta$ fulfilling $\beta=1+1/\alpha$, 
which yields values of $\beta$ close to 2.
The expression given by Eq. (\ref{lasegunda})
was in fact the first approach followed by Zipf's himself
\cite{Zanette_book}, and
is usually considered as equivalent 
to Eq. (\ref{laprimera})
\cite{Adamic_Huberman,Newman_05,Zanette_book,Li02,FontClos_Corral};
however, as it is derived in the continuum limit, 
both expressions can only be 
equivalent asymptotically, for large $n$ \cite{Mandelbrot61}.
Consequently, if one wants to be precise, a natural question follows:
which one is the ``true'' Zipf's law (if any)?

We cannot know a priori which of the two Zipf's laws better describes real texts, 
but we can argue which of the two representations (that of $n(r)$, Eq.\eqref{laprimera}, or that of $f(n)$, Eq.\eqref{lasegunda})
is better for statistical purposes,
independently of the functional dependency they provide.
It is clear that the rank-frequency representation, given by $n(r)$,
presents several difficulties, due to the peculiar nature of the rank variable.
First, in Ref. \cite{Corral_Cancho}, Zipf-like texts were randomly generated
following Eq. (\ref{laprimera}), keeping the original ranks ``hidden''
(as it happens in the real situation), and it was found that the rank reconstructed 
from the sample deviated considerably from the original ranks when 
these were taking large values
(which for a power law happens with a high probability). 
The resulting ML estimations of the exponent
$\alpha$ were highly biased and 
the Kolmogorov-Smirnov test rejected the power-law hypothesis, 
although the original ranks were power-law indeed.

One might argue that the problem could be escaped by using an upper truncated power-law distribution 
(introducing then an additional parameter for the truncation), 
in order to avoid the inconsistency of the rank representation for high values.
But a second problem is that
the rank is not a true random variable \cite{Kolmogorov}, 
as its values are assigned a posteriori,
once the sample (i.e., the text) is analyzed.
This means that the rank does not show ``enough'' statistical fluctuations, 
that is, if $r_a < r_b$, then the frequency of $a$ is always larger, by construction, than the frequency of $b$. 
This does not necessarily happen for a true random variable.
The negative correlation between the variable and its frequency of occurrence makes the 
{power-law hypothesis} harder to reject. 
In fact, inflated $p$-values (not uniformly distributed between 0 and 1) 
have been found when fitting truncated power laws to simulated power-law
rank-frequency representations \cite{Corral_Cancho}.
This problem could still be avoided by choosing a low enough upper truncation parameter
(yielding a very short range of ranks, for which the fluctuations would be very little) 
but at the expense of disregarding an important part of the data.

A third inconvenience is the impossibility, due to normalization, that a non-truncated power law
comprises values of the $\alpha-$exponent smaller than 1.
This yields the necessity of introducing a truncation parameter
that may be artifactual, i.e., not present in the real system.
All this leads to the conclusion that the most reliable method of parameter estimation 
(ML, in a frequentist framework) 
cannot be directly applied to the rank-frequency representation.
In contrast, the representation in terms of the distribution of frequencies
is devoid of these problems \cite{Corral_Cancho}, 
as $n$ is a well-defined random variable,
and this will be the representation used in this paper for statistical inference.
Nevertheless, for alternative arguments, see Ref. \cite{Altmann_Gerlach}.

The purpose of this paper is to quantify, at a large, big-data scale, 
different versions of Zipf's law and their ranges of validity. 
In the next section, we present and justify the three Zipf-like distributions
we are going to fit, 
and we briefly explain the selected fitting method and the goodness-of-fit test.
The corpus of texts under consideration is also detailed. 
The following section
presents the results, with special attention to their statistical significance
and their dependence with text length.
Finally, we end with the conclusions and some technical appendices.

\section{Zipf-like distributions}
\label{secdos}

As implicit in the introduction, and in contrast with continuous random variables, 
in the discrete case a power law in the probability mass function $f(n)$ does not
lead to a power law in the complementary cumulative distribution or survival function $S(n)$, 
and vice-versa.
Let us specify our definition for both functions,
$f(n)=\mbox{Prob}[\mbox{frequency} =n]$ (as usual),
and $S(n)=\mbox{Prob}[\mbox{frequency} \ge n]$
(changing, for convenience, the usual strict inequality sign by the non-strict inequality).
Then, the relation between both is $f(n)=S(n)-S(n+1)$ and
$S(n)=\sum_{n'=n}^\infty f(n')$.

We consider that the values the random variable takes, given by $n$, are discrete,
starting at the integer value $a$, 
taking values then 
$
n=a, a+1,\dots
$
up to infinity.
In this study we will fix the parameter $a$ to $a=1$,
in order to fit the whole distribution and not just the tail.
Then, although 
for large $n$ and smooth $S(n)$ we may approximate
$f(n) \simeq - dS(n) / dn$, 
this simplification, which lies 
in the equivalence between Eqs. (\ref{laprimera}) and (\ref{lasegunda}),
is clearly wrong for small $n$.

For the first distribution that we consider, the power-law form is in $f(n)$, then,
\begin{equation}
\label{distro1}
f_1(n)=\frac{1}{\zeta(\beta,a) n^{\beta}}.
\end{equation}
 This is just the normalized version of Eq. (\ref{lasegunda}),  and then,
$$
S_1(n)=\frac{\zeta(\beta,n)}{\zeta(\beta,a)}
$$
with $\beta > 1$ and
$\zeta(\beta,a)=\sum_{k=0}^{\infty}{(a+k)^{-\beta}}$ denotes the Hurwitz zeta function, 
which ensures normalization of both expressions.
A preliminary analysis in terms of this distribution was done
in Ref. \cite{Font_Clos_Corral}.
In contrast, when the power law is in $S(n)$, this leads to our second case,
\begin{equation}
\label{distro2}
f_2(n)=\left(\frac{a}{n}\right)^{\beta-1}-\left(\frac{a}{n+1}\right)^{\beta-1}
\end{equation}
and
$$
S_2(n)=\left(\frac{a}{n}\right)^{\beta-1} 
$$
with $\beta > 1$ again.
Note that this corresponds to a power law in the empirical rank-frequency relation.

Finally, it is interesting to consider also the frequency distribution 
derived by Mandelbrot \cite{Mandelbrot61} when ranks are generated independently from a power law in 
Eq. (\ref{laprimera}), which is,
\begin{equation}
\label{distro3}
f_3(n)=
\frac{(\beta-1)\Gamma(a)}{\Gamma(a+1-\beta)}
\frac{\Gamma(n+1-\beta)}{\Gamma(n+1)}
\end{equation}
and
$$
S_3(n)=\frac{\Gamma(a)\Gamma(n+1-\beta)}{\Gamma(n)\Gamma(a+1-\beta)},
$$
with $1 < \beta < 2$ and $\Gamma(\gamma)=\int_0^\infty x^{\gamma-1} e^{-x}dx$ denotes
the gamma function \cite{Abramowitz}.
In this case the power law is the underlying theoretical rank-frequency relation $n(r)$.
Note that $f_3(n)$ can be written as
$$
f_3(n) = \frac{B(n+1-\beta,\beta)}{B(a+1-\beta,\beta-1)}
$$
using the beta function \cite{Abramowitz}, $B(x,y)=\Gamma(x)\Gamma(y)/\Gamma(x+y)$,
with an analogous expression for $S_3(n)$,
but do not confuse this distribution with the beta distribution.

In all three cases it is easy to show that we have well-defined, normalized probability distributions, when $n$ takes values $n=a, a+1, \dots $, with $a$ being a positive integer.
Moreover, in the limit $n\rightarrow \infty$ all of them yield a power-law tail,
$f(n) \propto 1/n^\beta$,
 so $\beta$ will be referred to as the power-law exponent.
 Indeed, it is easy to show that
$$f_2(n) \xrightarrow{n\to\infty} (\beta-1)
\frac{a^{\beta-1}}{n^\beta},$$ 
whereas 
$$
f_3(n) \xrightarrow{n\to\infty} \frac{(\beta-1)\Gamma(a)}{\Gamma(a+1-\beta) n^\beta}
$$
using Stirling's formula \cite{Abramowitz}.
The main difference between the three distributions is 
in the smaller values of $n$, taking $f_2(n)$ a convex shape
in log-scale (as seen ``from above'');
$f_3(n)$ a concave one;
and $f_1(n)$ being somehow in between, as it is neither concave nor convex.

\section{Methodology and data}
In order to fit these three distributions to the different texts, 
and test the goodness of such fits, we use maximum likelihood estimation \cite{Pawitan2001}
followed by the Kolmogorov-Smirnov (KS) test \cite{Press_C}.
The procedure seems 
similar to the one proposed in Ref. \cite{Clauset},
but as $a$ is fixed here, the problems resulting from the search of the optimum $a$ \cite{Corral_nuclear,Corral_Deluca} do not arise in this case.

The method of ML estimation proceeds in the following simple way.
Given a set of data $\{n_i\}$ with $i=1,2, \dots N$, 
and a probability mass function parameterized by $\beta$, denoted as 
$f(n;\beta)=f(n)$,
the log-likelihood function is obtained as
\begin{equation}
l(\beta) = \sum_{i=1}^N \ln f(n_i;\beta).
\label{loglikelihood}
\end{equation}
We are assuming that the data points $n_i$ are independent from each other, 
in other words, we are calculating the likelihood that the data are
generated independently from $f(n;\beta)$.
The ML estimation of $\beta$ is obtained as the value of $\beta$
which maximizes $l(\beta)$; we undertake this numerically, 
using Brent's method \cite{Press_C}.  
In the case of the distribution $f_1$ the log-likelihood function takes the simple form 
$l_1(\beta)/N=-\ln(\zeta(\beta,a)) -\beta \ln G$, with $G$ the geometric mean
of the set $\{n_i\}$.
For the other distributions no closed-form expression
is possible and we use Eq. (\ref{loglikelihood}) directly.

As mentioned, the goodness-of-fit test is done through the Kolmogorov-Smirnov
statistic \cite{Press_C,Clauset}, in the discrete case \cite{Corral_Deluca_arxiv}, 
for which the $p$-value is calculated from Monte Carlo simulations
(due to the fact that, as the value of the exponent 
is calculated from the same data is going to be tested, 
the theoretically computed $p$-value \cite{Press_C}
would be positively biased).
In this paper we use 100 Monte Carlo simulations for each test.
The proper simulation of the 3 distributions is explained in the Appendix.
 Remember that a small enough $p$-value
leads to the rejection of the fit.
Although we perform multiple testing, we do not incorporate any Bonferroni-like correction
\cite{Abdi_Bonferroni,Bland_Altman,Benjamini}, due to the fact that 
these corrections increase the number of non-rejected null hypothesis
(that is, decrease the number of type I errors),
inflating the performance of the fits, in the case of goodness-of-fit tests. 
Without Bonferroni-like corrections, our acceptance (i.e., non-rejection) 
of the fits is more strict.

In order to apply this methodology we consider a set of 37\,041 texts 
stored in UTF-8 encoding in the Project Gutenberg database (accessed July 2014 \cite{Gutenberg}).
These texts correspond to different languages, styles, 
and time periods, although
most of them are works of literature from the Western cultural tradition \cite{EWikip}.
First of all, parts of text that do not pertain to the piece under consideration 
(copyright notes, headers,...) are removed by an automatized process. 
Books that have not been filtered in this step (mainly because they do not have standard delimiters) are discarded. 
After this, we still keep 97.5\,\% of the total (i.e., 36\,108). 
To perform our study, we restrict ourselves to the subset of texts in English, 
which represent the 86\,\% of these 36\,108 (i.e., 31\,102).
 
\begin{figure} [!htbp]
\begin{center}
\includegraphics[width=90mm]{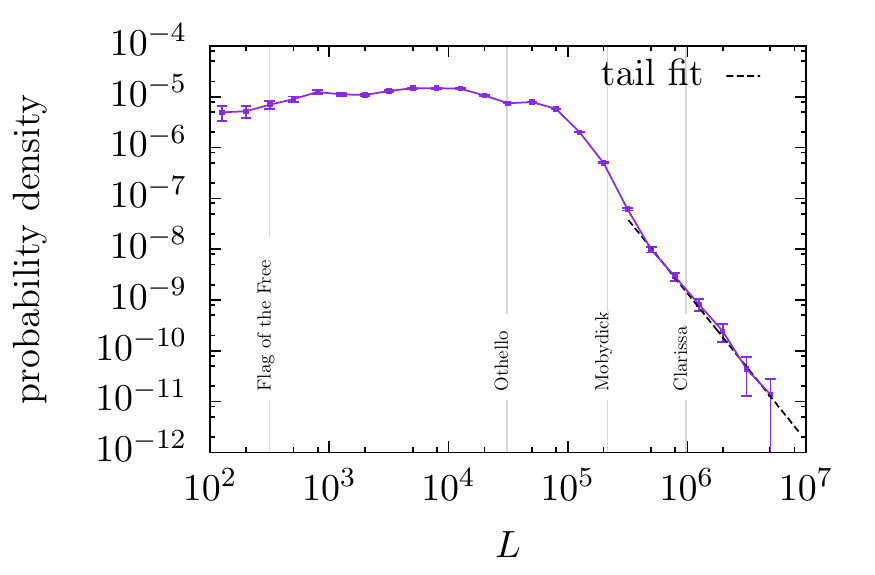}
\end{center}
\caption{
Estimation of the probability density function of text length $L$ 
in the English Project Gutenberg database,
using logarithmic binning (5 bins per decade).
Texts with less than 100 tokens are not considered.
A power-law fit of the tail \cite{Corral_Deluca} yields an exponent $2.92 \pm 0.15$. 
% xmin=323593, p=0.24, power law continua
}
\label{Ldens}
\end{figure}

An important characteristic of each text  
is its length, $L$, counting the number of word tokens it contains.
It turns out to be that in the database $L$ expands from very small values up to 4\,659\,068 tokens, 
 with a distribution that is shown in
Fig.~\ref{Ldens}. Observe the roughly uniform distribution up to about $L=10^5$, 
and the decay afterwards.
We consider only the 31\,075 English texts that 
consist of more than 100 word tokens, 
 as smaller texts would not have statistical value.
For each of these texts 
we select then actual word types (punctuation signs, numbers and any character different from letters are not considered) to count their frequencies $n$,
which will be our primary object of study. 

The values of these frequencies, for each text, are available
on  http://dx.doi.org/10.6084/m9.figshare.1515919, in order to facilitate the reproducibility
of our results.
 
In summary, we apply the above described fitting and goodness-of-fit procedure
--using ML estimation and the Kolmogorov-Smirnov test-- to a total of 31\,075
texts from the English Project Gutenberg database, using three different possibilities
for the distribution of frequencies: $f_1$ (Eq.~\eqref{distro1}), $f_2$ (Eq.~\eqref{distro2}),
and $f_3$ (Eq.~\eqref{distro3}). This yields a total of $3\times31\,075$ 
fits and associated $p$-values, which we analyze and interpret in what follows.

\section{Results}\label{sectionresults}
\begin{figure}[!htbp]
\begin{center}
\includegraphics[width=90mm]{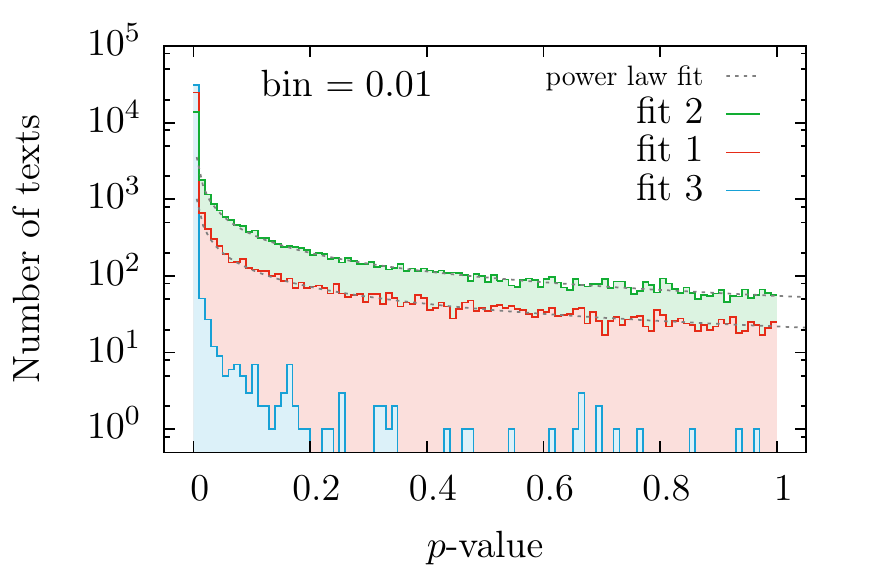}
\end{center}
\caption{
Histograms of $p$-values obtained
when the Zipf-like distributions $f_1$, $f_2$, and $f_3$
are fitted to the texts of the English Project Gutenberg.
The histograms just count the number of texts in each bin of width 0.01.
Note the poor performance of distribution 3
and the best performance of 2.
Power-law approximations to the histograms for $f_1$ and $f_2$,
with respective exponents 0.74 and 0.78,
are shown as a guide to the eye.
}
\label{pvfits_en}
\end{figure}

\begin{figure}[!htbp]
\begin{center}
\includegraphics[width=90mm]{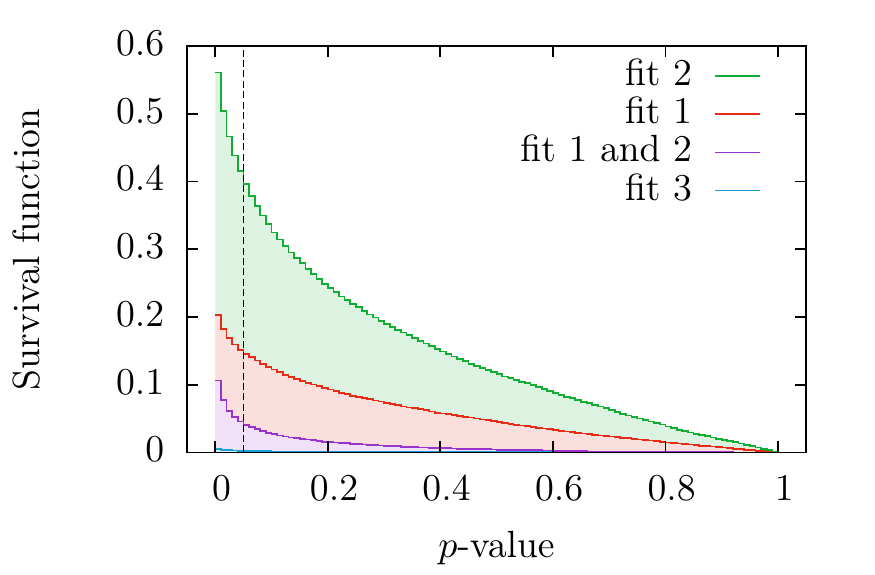}
\end{center}
\caption{
Complementary cumulative distributions 
(i.e., survival functions) of $p$-values
obtained when our three distributions
are fitted to the texts of the English Project Gutenberg. 
 This corresponds, except for normalization, to the integral of the previous figure,
but we have included a fourth curve for the fraction of texts whose $p$-values for fits $1$ and $2$
are both higher than the value marked in the abscissa.
Note that the values of $p$ can play the role of the significance level.
The value for $p=0$ is not shown, in order to have higher resolution. 
}
\label{Nvspv1y2}
\end{figure}

Contrary to previous studies where the number of texts considered was, at most, in the order of tens, the large-scale approach taken in this work requires a statistical analysis of the fitting results, as a case-by-case interpretation is out of hand.
We first focus on the distribution of $p$-values,
see Fig.~\ref{pvfits_en} and Fig.~\ref{Nvspv1y2}.
If \emph{all} texts were truly
generated by a mechanism following a given distribution, 
the corresponding $p$-values for that distribution 
would be uniformly distributed between zero and one. 
As seen in Fig. \ref{pvfits_en}, this is not the case and, furthermore, 
most texts have rather small $p$-values for the three fits;
nevertheless, for distributions $f_1$ and $f_2$
there are still many texts that yield high enough $p$-values. 
This implies that, although we cannot conclude that the whole database is  generated by any of these distributions, 
these cannot be rejected as good 
 descriptions for large subsets of the database. 
Regarding distribution $f_3$,
it is clear from the histogram of $p$-values that it can be discarded as a good description of 
the distribution of frequencies in any non-negligible subset of texts.
So, from now on, we will concentrate on the remaining options, $f_1$ and $f_2$,
to eventually quantify which of these better describes
our corpus.
In essence, what we are interested in is which version of Zipf's law, 
either distribution $f_1$ or $f_2$, fits better a reasonable number of texts, 
and which range of validity these simple one-parameter distributions have.

\begin{figure}[!htbp]
\begin{center}
\includegraphics[width=90mm]{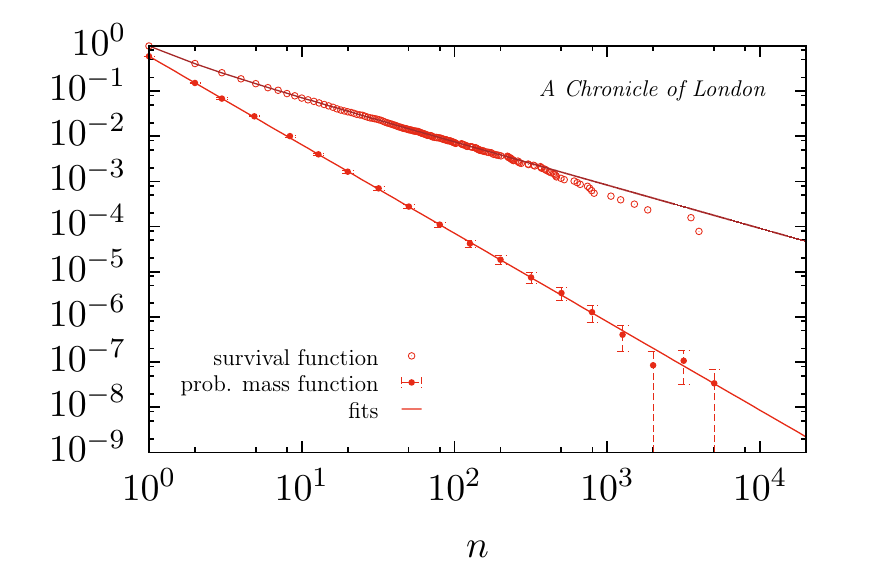}
\includegraphics[width=90mm]{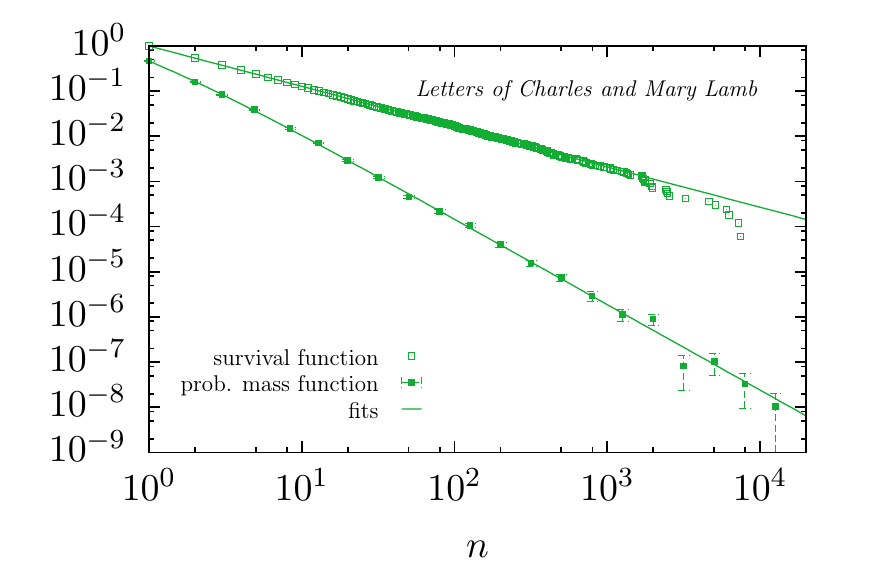}
\includegraphics[width=90mm]{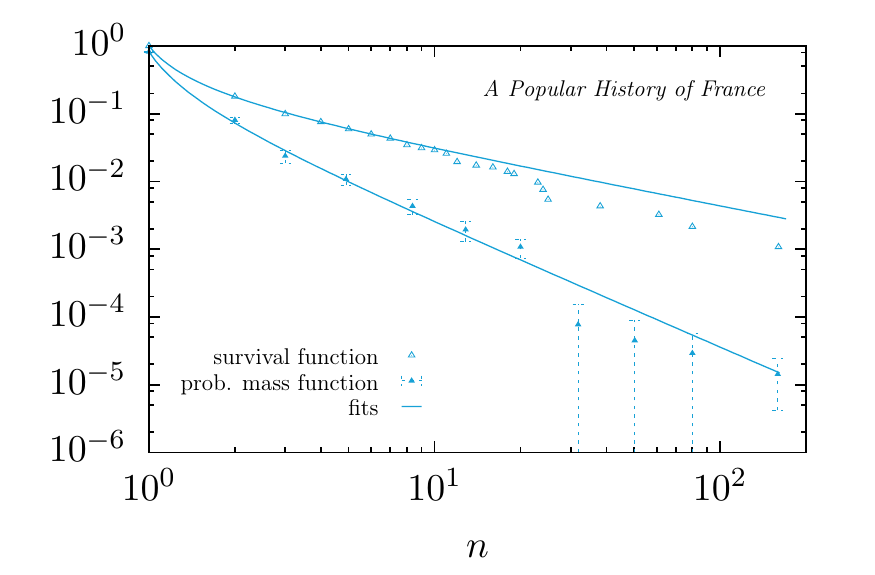}
\end{center}
\caption{
Complementary cumulative distribution and probability mass function of texts:
(a) \textit{A Chronicle of London, from 1089 to 1483};
(b) \textit{The letters of Charles and Mary Lamb}, edited by E. V. Lucas;
(c) \textit{A Popular History of France from the Earliest Times, Vol. I,} by F. Guizot.
These texts are the ones with the largest length $L$ (83\,720, 239\,018 and 2\,081  respectively)
of those that fulfill  $p> 1/2$, for fits 1, 2 and 3 
respectively.
The exponent $\beta$ takes values 1.96, 1.89, and 1.82,
in each case. 
}
\label{visual}
\end{figure}

The outcome is that, independently of the significance level
(as long as this 
is not below our resolution of 0.01 given the number of Monte Carlo simulations),
the ratio between the number of texts fitted by distribution $f_2$
and those fitted by $f_1$ is nearly constant, taking a value around 2.6.
For example, considering significance level (i.e., minimum $p$-value)
equal to 0.05,
Fig. \ref{Nvspv1y2} shows that 
distribution $f_2$ fits about 40\,\% of all texts, 
whereas distribution $f_1$ fits just 15\,\%.
Both percentages include a 2.7\,\%
of texts that are fitted by both distributions simultaneously, although 
this number does not keep a constant ratio with the other two, rather, it
decreases when the significance level is increased 
(as it is implicit in the values of Fig. \ref{Nvspv1y2}).  
Given that the aforementioned ratio of 2.6 is independent of the significance level,
it is fair to say that distribution $f_2$ provides, compared to $f_1$, a better description of our database.
As a visual illustration of the performance of the fits we
display in Fig.~\ref{visual} the word frequency distribution of
the longest texts that have $p > 0.5$, for distributions  $f_1$, $f_2$ and $f_3$. 

\begin{figure} [!htbp]
\begin{center}
\includegraphics[width=90mm]{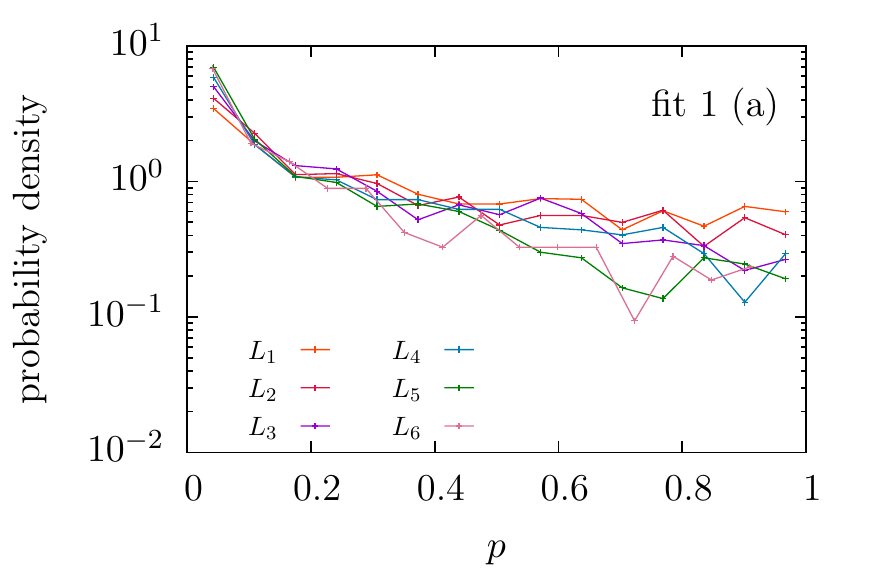}
\includegraphics[width=90mm]{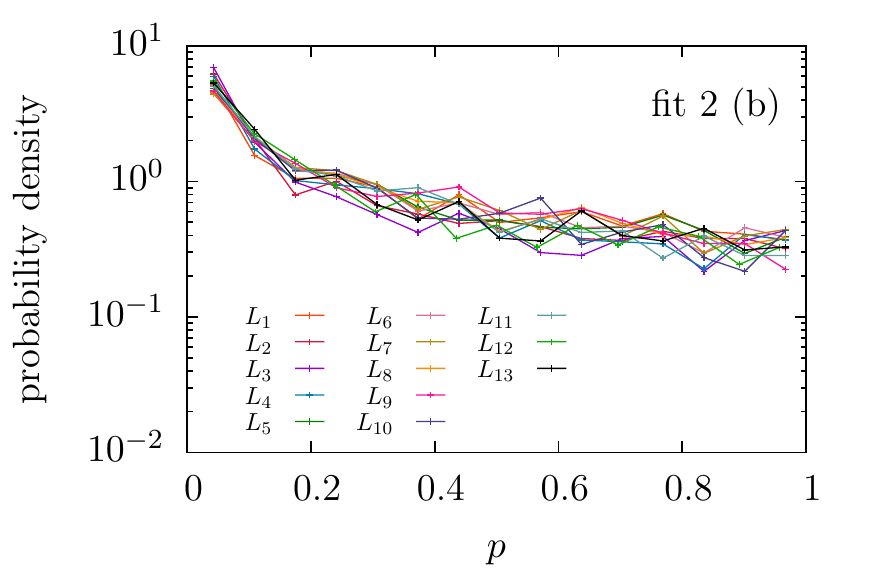}
\end{center}
\caption{
Estimated probability  density functions 
of $p$-values conditioned to $p\ge 0.01$ 
separating for different ranges of text length $L$. $p$-values correspond
to the fitting of  word frequencies to (a) distribution $f_1$ and (b) distribution $f_2$.
We divide the distribution of text length into 15 intervals of 2\,000 texts each. 
For distribution $f_1$ only the first seven groups (up to length 34\,400) are displayed (beyond this value we do not have enough statistics to see the distribution of  $p$-values greater than 0.01, as displayed in Fig.~\ref{pv0};
for distribution 2  this happens only in the last two groups). The intervals $L_i$ range from $L_{1}=[115, 5\,291]$ to $L_{13}=[89\,476, 103\,767]$.
}
\label{pv_Ls_fit2}
\end{figure}

\begin{figure}[!htbp]
\begin{center}
\includegraphics[width=90mm]{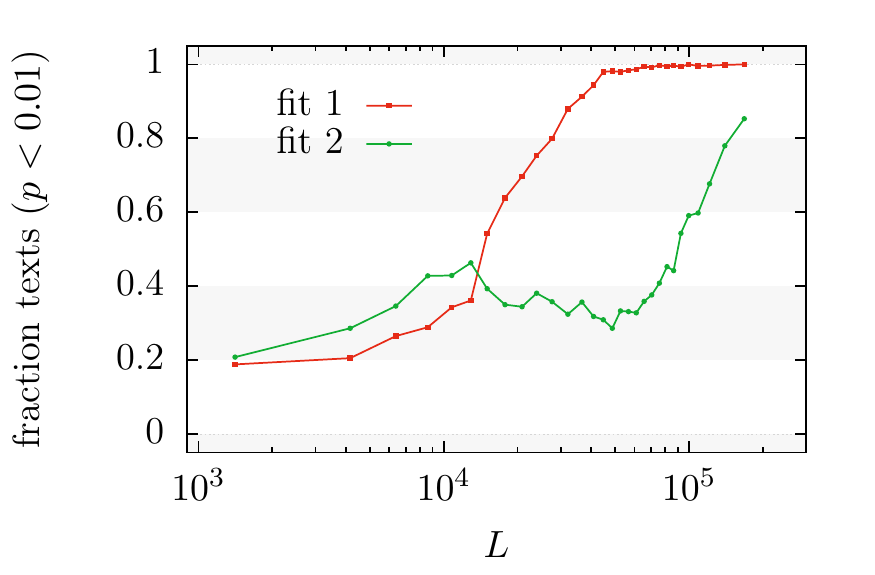} 
\end{center}
\caption{Number of texts with $p$-value near zero ($p<0.01$) 
in different ranges of $L$ divided by the number of texts in the same ranges,
for the fits of distributions $f_1$ and $f_2$. 
Values of $L$ denote the geometric mean of ranges containing 1000 texts each.
The higher value for fit 1 (except for $L$ below about 13000 tokens) denotes its worst performance. 
}
\label{pv0}
\end{figure}

The next question we address is the dependence of the performance of  fits on text length $L$.
In order to  asses this,
note that from the shape of the histograms in Fig.~\ref{pvfits_en} we can distinguish two groups of texts:
those that lie in the zero bin (whose $p$-value is strictly less than 0.01), and the rest. 
Taking the last group, i.e., texts with $p\geq 0.01$,
and partitioning it into different subsets according to text length
(i.e., looking at the distribution of $p$ conditioned to $p\ge 0.01$ for different ranges of $L$), 
it holds that the shape of the resulting distribution of $p$ does not strongly depend on $L$, as shown in Fig.~\ref{pv_Ls_fit2}.
In contrast, the number of texts that yield $p$-value near zero certainly varies with $L$, see Fig.~\ref{pv0}.
Therefore, in order to compare the performances of $f_1$ and $f_2$ as a function of the text length $L$,
it is enough to consider a single value of the significance level (greater than zero)
as the results for any other significance level 
will be  the same, in relative terms.

\begin{figure}[!htbp]
\begin{center}
\includegraphics[width=90mm]{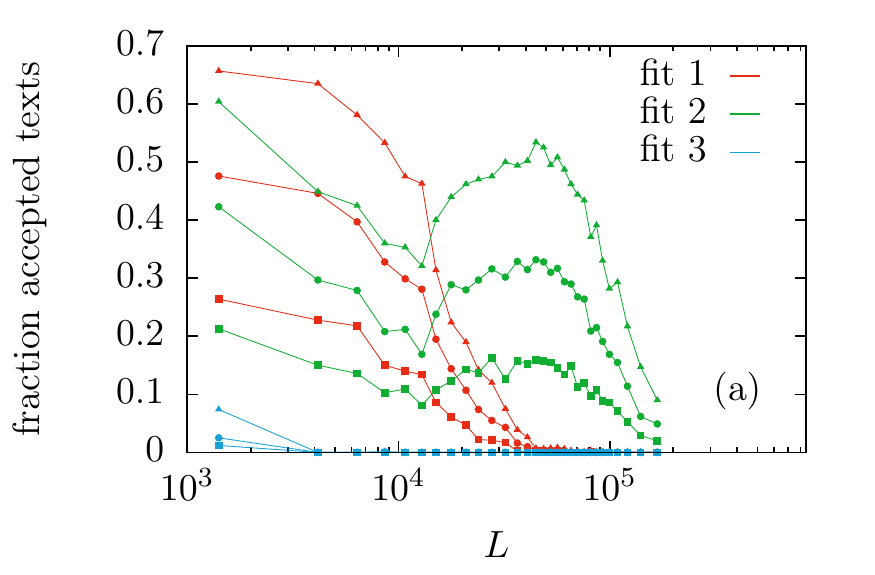}
\includegraphics[width=90mm]{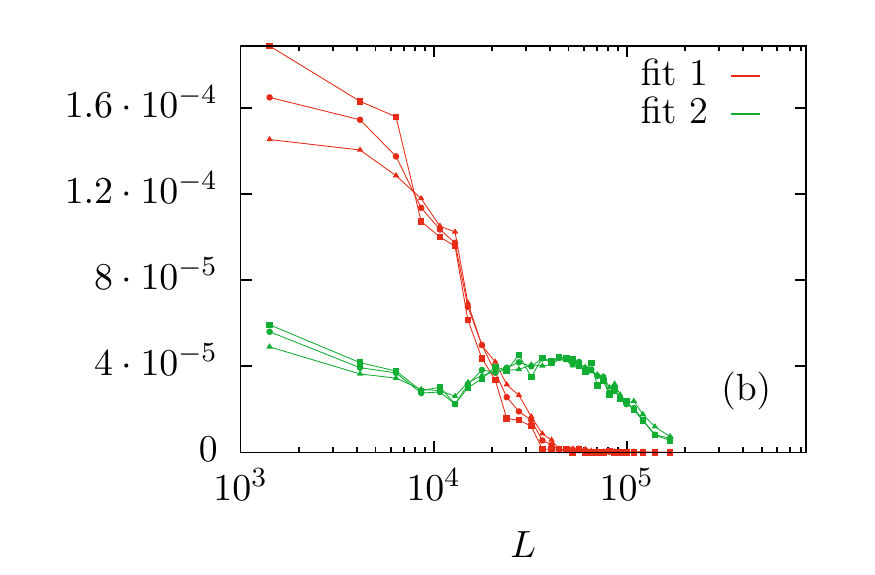}
\end{center}
\caption{
(a)  Histograms showing the fraction of accepted texts by the three distributions
as a function of their text length, for three different significance levels $p_0$: 
0.05 (upper curves), 0.20 (middle), 0.50 (lower).
To be concrete, for each range of $L$, 
the ratio between the number of texts with $p\ge p_0$
and the number of texts in that range is calculated.
(b) Same curves (removing those for distribution 3) under rescaling.
We rescale  the $y-$axis by the number of $p\ge p_0$, in each case,
showing that the relative performance of each fit
with regard $L$ is independent on the significance level.
Bins are selected to contain 1000 texts each.
}
\label{perc_L}
\end{figure}

Indeed, Fig.~\ref{perc_L}(a) shows how distribution $f_1$ fits some more texts than
distribution $f_2$ for small values of $L$, up to about $13\,000$ tokens.
But for larger texts, distribution $f_2$ clearly outperforms distribution $f_1$,
which becomes irrelevant for $L$ beyond $100\,000$ (at 0.05 significance level), 
whereas distribution $f_2$ is able to fit many texts with $L$ larger than $200\,000$.
The figure shows that this is the case no matter if the significance level is 0.05, 0.20, or 0.50;
the collapse of the curves in  Fig.~\ref{perc_L}(b) confirms this fact.
From Fig.~\ref{pv0} one could infer the same for significance level equal to 0.01.
This stability of the performance of the fits for different significance levels arises from 
the observed fact that the distributions of $p$-values (conditioned to $p \ge 0.01$)
are nearly identical for different $L$,
as shown in Fig.~\ref{pv_Ls_fit2}.
      
In order to  check the consistency of our fitting procedure, we also perform a direct comparison of models through the likelihood ratio (LR) test \cite{Vuong,Clauset}.
We apply this test to all texts that have been fitted, 
considering 0.05 as significance level,
by at least one of the two distributions $f_1$ and $f_2$.
Then the log-likelihood-ratio between distributions $f_1$ and $f_2$ is 
\begin{equation}
R_{1,2}= \sum_{i=1}^{N}\left(\ln f_1(n_i)-\ln f_2(n_i)\right), \nonumber
\end{equation}
and, under the null hypothesis that both models are equally good to describe the data,
$R_{1,2}$ should be normally distributed with zero mean and a variance that can be estimated as $N \sigma^2$,
with $\sigma^2$ the variance of the random variable
$\ln f_1(n)-\ln f_2(n)$.
Large absolute values of $R_{1,2}$ will lead to the rejection of the null hypothesis.

Table \ref{tabla_lkh} merges the results of the LR test and our previous procedure (based on 
ML estimation plus the KS test). 
The total number of texts previously fitted by $f_1$ or/and $f_2$ is displayed depending on the sign of the corresponding log-ratio $R_{1,2}$.
However we must take into account that the sign of the obtained value of $R_{1,2}$ could be a product of just statistical fluctuations if the true value were zero and thus, the sign of $R_{1,2}$ 
cannot be trusted in order to discriminate between two models.
The probability under the null hypothesis, of obtaining an absolute value of the log-ratio greater than the empirical value
$|R_{1,2}|$ is computed through:  
\begin{equation}
p_{LR}=
\mbox{erfc}\left(\frac{|R_{1,2}|}{\sqrt{2N\sigma^2}}\right)\nonumber
\end{equation}
where 
erfc  is the complementary error function \cite{Abramowitz}.
We will take as statistically significant those $R_{1,2}$ that yield $p_{LR}<0.05$.
Equivalently, at 0.05 significance level,
$R_{1,2}$ is significant if its absolute value is greater than 
$R_c=1.96 \sqrt{N \sigma^2}$. The results are shown in Table \ref{tabla_lkh_pv}

Note that the LR test cannot conclude if a fit is good or bad, 
as it only compares the relative performance of two fits;
in other words, if the LR test selects a particular distribution, 
that distribution can still yield a bad fit, in absolute terms.
Anyway, there is no mismatch between the results of both tests: 
any time the ML-KS method selects one distribution over the other,
the LR test either supports the selection or does not give significant results, 
but it never selects the other option (as shown in Table~\ref{tabla_lkh_pv}).

\begin{center}
\begin{table}[!htbp]  
\begin{tabular}{| l | r | r | r |}\hline
                    & $R_{1,2}>0$ & $R_{1,2}<0$ & Total ML-KS \\\hline
$f_1$ (exclusively) & 3614        & 81          & 3695        \\\hline
$f_2$ (exclusively) & 120         & 11366       & 11486       \\\hline 
$f_1$ and $f_2$     & 431         & 398         & 829         \\\hline 
Total LR            & 4165        & 11845       & 16010       \\\hline 
\end{tabular}
\caption[]{
The number of texts that are fitted by $f_1$ or $f_2$ or both
at 0.05 significance level of the ML-KS procedure,
separated into two columns according to the sign of $R_{1,2}$. 
Positive $R_{1,2}$ means that the likelihood 
for $f_1$ is greater than that for $f_2$,
and conversely for negative $R_{1,2}$.
Nevertheless, the sign of $R_{1,2}$ is not an indication of significance,
for significant LR tests see
Table~\ref{tabla_lkh_pv}.
}
\label{tabla_lkh}
\end{table}
\end{center}

\begin{center}
\begin{table}[!htbp] 
\begin{tabular}{| l | r | r |}\hline
                                & $R_{1,2}>R_c$ & $R_{1,2}<-R_c$  \\\hline
$f_1$ (exclusively)             & 1666          & 0               \\\hline
$f_2$ (exclusively)             & 0             & 9423            \\\hline 
$f_1$ and $f_2$                 & 0             & 3               \\\hline 
Total LR test                   & 1666          & 9426            \\\hline 
None (neither $f_1$ nor $f_2$)  & 510           & 11431           \\\hline 
\end{tabular}
\caption[]{
Number of texts with a significant LR test, 
at the 0.05 level, either favouring distribution $f_1$ 
($R_{1,2}>R_c$)
or distribution $f_2$ $(R_{1,2}<-R_c)$,
for different outcomes of the ML-KS procedure (at the 0.05 level also). 
Note that these cases correspond to a subset of the previous table.
An additional row shows the number of texts that are
fitted neither by distribution $f_1$ nor $f_2$;
notice that in this case a significant LR test does not guarantee a good fit.
}
\label{tabla_lkh_pv}
\end{table}
\end{center}

\begin{figure}[!htbp]
\begin{center}
\includegraphics[width=90mm]{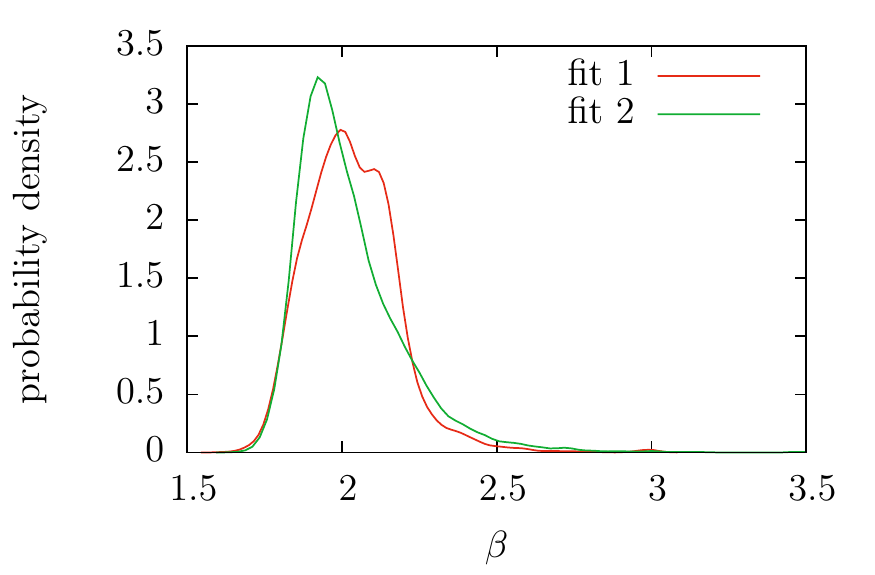}
\end{center}
\caption{
Estimation of the probability density  of the exponent $\beta$ 
for texts yielding $p\ge0.05$ in fit 1 and fit 2.
Curves have been calculated from the histograms via normal kernel smoothing method
as implemented in MatLab (\textit{ksdensity} function).
Estimated mean and standard deviation are 2.03 and 0.15 respectively for fit 1, and 2.02 and 0.17 for fit 2.}
\label{betadist}
\end{figure}
      
\begin{figure}[!htbp]
\begin{center}
\includegraphics[width=90mm]{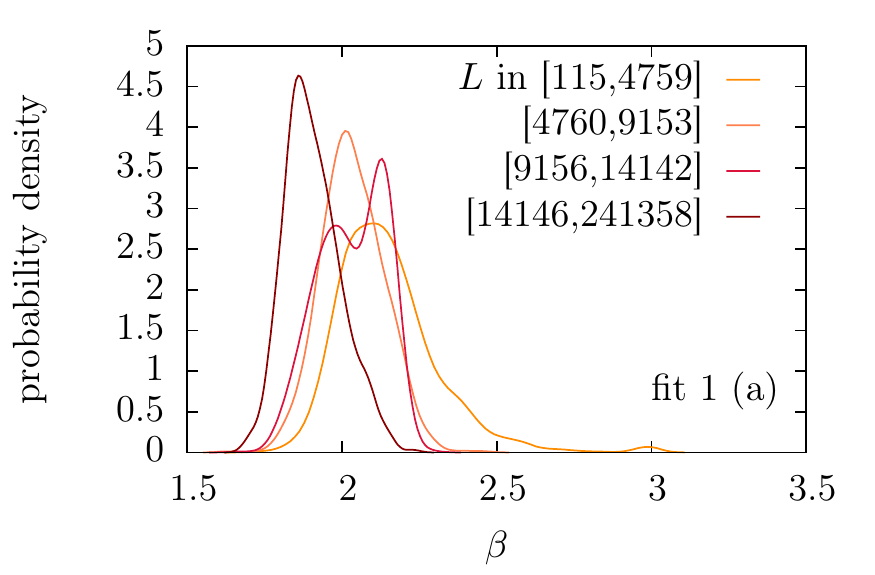}
\includegraphics[width=90mm]{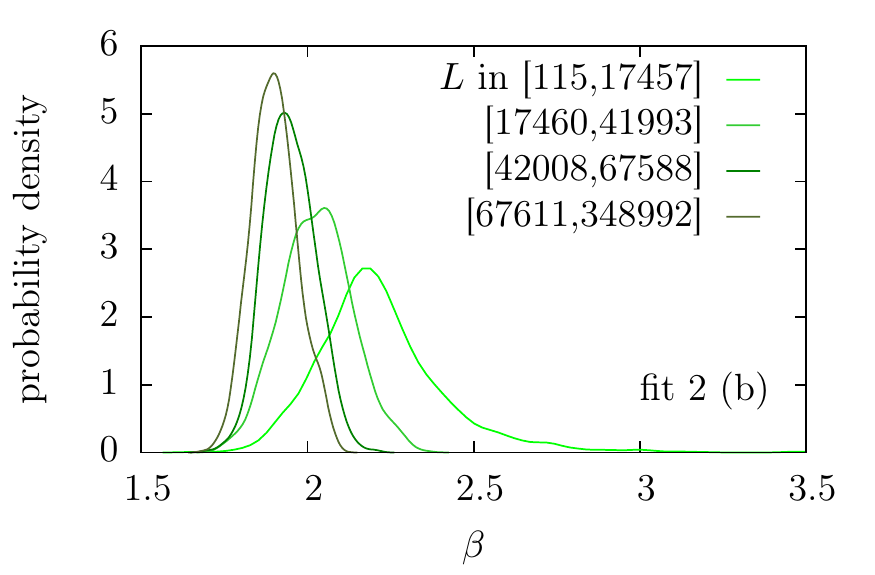}
\end{center}
\caption{
Estimated probability density  of $\beta$ for fits with $p\ge 0.05$, in different length ranges.  
We have divided both groups of accepted texts into 4 percentiles according to $L$.
As in the previous figure, the normal kernel smoothing method is applied.
(a) For distribution $f_1$.
(b) For distribution $f_2$.
}
\label{betadist_L}
\end{figure}

Taking now those texts whose frequency distributions could be approximated by $f_1$ or $f_2$, we draw attention to the distribution of the estimated exponents (i.e., the parameter $\beta$). 
The original formulation of Zipf's law implies $\beta=2$ and 
Fig.~\ref{betadist} shows that $\beta$ is certainly distributed around 2, 
with a  bell-like shape.
If we check the effect of the text length $L$ in the distribution of $\beta$, 
we find a decreasing trend of $\beta$ with $L$, as can be seen in Fig.~\ref{betadist_L}.
We have tested that this observation is not an artifact of the fitting method, as synthetic texts generated with fixed $\beta$ do not show this behavior.
We have no theoretical explanation for this fact, but notice that this trend
is not in disagreement with the claims of Ref. \cite{FontClos_Corral}, 
where the stability of the exponent $\beta$ was demonstrated
for a single growing text (i.e., comparing small parts of a text with the whole).

\section{Conclusions}
Zipf's law is probably the most intriguing and at the same time well-studied experimental law of quantitative linguistics, and extremely popular in its wider sense in the science of complex systems.
Although the previous literature is vast, as far as we know our work constitutes the first large-scale
analysis of Zipf's law in single (non-aggregated) texts. 
Thus, we are in a position to make a well-grounded statement about the validity of Zipf's law in such texts.

Let us first briefly summarize, however, some key technical points of our study.
First, we have analyzed a total of 31\,075 English texts from the Project Gutenberg database using rigorous fitting procedures, 
and have tested how well they are described by three Zipf-like distributions.
Our choice of distributions has not been exhaustive;
rather, we have limited ourselves to different interpretations 
of what can be understood as ``Zipf's law'', 
in the sense of having a perfect power law either in the probability mass function
of word frequencies, 
or in the complementary cumulative distribution function
(whose empirical estimation leads to the rank-frequency relation of the sample),
or in the rank-frequency relation of an underlying population.
Remarkably, the resulting distributions have a unique parameter, $\beta$, 
which in all cases is the exponent of an asymptotic power law 
in the probability mass function of the frequency.
It is left to explore how other, more complicated extensions of Zipf's law perform on this large corpus, but it is obvious that, by including additional parameters, one might provide good fits to a larger number of texts
(although in this case, proper model selection will require to balance number of parameters and parsimony).

Our aim in this paper has not been to fit as many texts as possible, but to test the
performance of the simplest Zipf-like distributions within a very strict, conservative framework.
Indeed, by requiring the three versions of Zipf's law to hold
on the full range of frequencies $n=1,2,\dots$ (and not only on the tail of the distribution)
we put ourselves in the strictest range of demands.
It is hence remarkable that, e.g., at the standard significance level of $0.05$, 
and for text lengths between $10^4$ and $10^5$
word tokens, more than 40\,\% of the considered texts are statistically compatible 
with the pure power law in the 
complementary cumulative distribution function represented
by distribution $f_2$ (see Fig.~\ref{perc_L}).
So, we can state that, for the corpus under consideration, 
the most appropriate version of Zipf's law
is given by a probability mass function
$$
f(n) = \mbox{Prob}[\mbox{frequency} = n] = 
\frac 1 {n^{\beta-1}} - \frac 1 {(n+1)^{\beta-1}},
$$
or, equivalently, by a complementary cumulative distribution function
$$
S(n) = \mbox{Prob}[\mbox{frequency} \ge n]=
\frac 1 {n^{\beta-1}}.
$$
Due to the broad coverage of the Project Gutenberg corpus
we speculate that this distribution should fit
a large fraction of generic (non-technical) English texts. 
Of course, testing this speculation in front of all possible corpora
is an impossible task.

We have also shown that our conclusions regarding the \emph{relative} performance of 
a pure power law in the probability mass function, given by
distribution $f_1$, versus distribution $f_2$ are robust with respect to changes in the significance level: 
about twice as many texts are statistically compatible with distribution $f_2$ than
those compatible with $f_1$, at any significance level  
(obviously, in absolute terms, the number of accepted texts varies with the significance level).
Hence we can conclude that
distribution $f_2$ gives a better description of English texts than distribution $f_1$,
at least for the corpus considered in this work.
We also conclude that distribution $f_3$, first derived by Mandelbrot \cite{Mandelbrot61},
is irrelevant for the description of texts in this corpus.
Finally, we have corroborated that the exponent $\beta$ of Zipf's law certainly varies from text to text, as had been previously claimed using other approaches for defining 
what Zipf's law is \cite{Zanette_2005,Corral_Boleda}.
Interestingly, the value $\beta=2$ originally proposed by Zipf himself is among the most frequent ones.

We believe that our analysis constitutes a major advancement in the understanding of Zipf's law.
It is astonishing how good the simplest one-parameter Zipf-like distributions perform on such a large set of texts, particularly with the strict set of requirements we have imposed.
This is in sharp contrast for instance with Zipf's law in demography \cite{Malevergne_Sornette_umpu} 
and in the distribution of income \cite{Drgulescu2001}, 
where the power law seems to be valid only for the tail corresponding to the largest sizes,
as it happens also for the distribution of word frequency in large text corpora,
as mentioned above \cite{Baroni2009,Ferrer2000a,Petersen_scirep,Gerlach_Altmann,Williams}.

Zipf's law has been subject to much debate, and will probably continue to be so for many years.
Indeed, one can always cast doubt on its validity on the basis of some particular examples.
Yet it seems clear to us that, in our modern times of big data and large computational capabilities, more efforts should be put towards large-scale analysis of Zipf's law.
We hope this paper constitutes a first step in this direction.

\section{Appendix: Simulation of discrete Zipf-like distributions}
As part of the testing procedure, we need simulated samples from $f_1$, $f_2$, and $f_3$,
which are discrete distributions defined for $n=a, a+1, \dots$. 
We will give the recipe of simulation for an arbitrary positive integer value of the lower cut-off $a$. 
It is simpler to start with $f_2$, as this is used as an auxiliary distribution
in the simulation of the other two. 
\\
{\it Simulation of $f_2$.}
Fixed $a$ and given the parameter $\beta$,  we want a set of random numbers  whose cumulative distribution function is a discrete power law: 
$S_2(n)=\left({a}/{n}\right)^{\beta -1}$. 
For that, we first generate a random number $u$ from a uniform distribution in the interval 
$(0, u_{max})$, with $u_{max}={1}/{a^{\beta-1}}$.
If we take $x={1}/{u^{1/(\beta -1)}}$, it turns out to be that the values of  $x$ yield a continuous power law with $S^c_2(x)=\left({a}/{x}\right)^{\beta -1}$, for $x\ge a$, where the superscript $c$ distinguishes the continuous distribution from its discrete analogous one.
So, taking $n$ equal to the integer part of $x$, i.e., 
$n=\text{int}(x)$, yields a discrete distribution with $S_2(n)=\left({a}/{n}\right)^{\beta -1}$, as desired.
This is so because, for any $X$, $\text{int}(X)\ge n$ is equivalent to $X\ge n$ for $n$ integer.
In a recipe:
\begin{itemize}
\item generate $u$ from a uniform distribution in $(0,1/a^{\beta-1}]$,
\item calculate $x= {1}/{u^{1/(\beta -1)}}$,
\item take $n =\text{int}(x)$.
\end{itemize}

By means of the so-called rejection method  \cite{Devroye},
simulated integers distributed following $f_2$ can be used for the simulation of
integers  following $f_1$ or $f_3$.
The key point to achieve a high performance in the rejection method is to use a ``good'' auxiliary function,
i.e., one that leads to a low rejection rate.
This is certainly the case in our framework, as explained below.
\\
{\it Simulation of $f_1$.}
In this case, the steps are:
\begin{itemize}
\item generate $n$ from $f_2(n)$,
\item  generate  $v$ from a uniform distribution in the unit interval,
\item  $n$ is accepted if 
$$v\le\frac{f_1(n)}{f_2(n)C},$$
and rejected otherwise,
where $C$ is the rejection constant given by 
$C=\max_{n\ge a}\left\{{f_1(n)}/{f_2(n)}\right\}$.
\end{itemize}

It is easy to check that the maximum of $f_1/f_2$ is reached at $n=a$ as this is a decreasing function
\cite{Corral_Cancho}.
The acceptance condition above can be simplified by
taking $\tau=(1+n^{-1})^{\beta-1}$, and $b=(1+a^{-1})^{\beta-1}$, 
then, the condition becomes: 
$$
b v n (\tau-1) \le a (b-1) \tau,
$$
which is devoid of the calculation of the Hurwitz-zeta function.
This is a generalization for $a>1$ of the method of Ref. \cite{Devroye}.
The choice of $f_2$ as the auxiliary distribution function is justified by the small value that $C$ takes, as this is the expected number of  generated values of $n$ until we accept one. 
For instance, for $\beta=2$ and $a=1$ we get $C=1.2158$.
\\
{\it Simulation of $f_3$.}
Proceeding similarly, we get in this case low values of 
$C=\max_{n\ge a}\left\{{f_3(n)}/{f_2(n)}\right\}$ as well 
(we get $C=2$ in the limit $\beta\rightarrow 2$ for $a=1$).
The maximum of ${f_3(n)}/{f_2(n)}$ is numerically seen to be reached at $n=a$.
In summary, the steps are:
\begin{itemize}
\item generate $n$ from $f_2(n)$
\item  generate  $v$ from a uniform distribution in the unit interval
 \item $n$ is accepted if \begin{equation}
 vf_2(n)\le a\left(1-\left(\frac{a}{a+1}\right)^{\beta-1}\right)\frac{\Gamma(n-(\beta-1))}{\Gamma(n+1)}\frac{\Gamma(a)}{\Gamma(1+a-\beta)}\nonumber
\end{equation}
and rejected otherwise.
\end{itemize}

\section{Acknowledgments}
We are grateful to the Project Gutenberg initiative, 
and to those who help maintain it alive.
S. Pueyo provided bibliographic wisdom for Zipf's law in ecology,
and I. Serra assistance for ML estimation.
I. M.-S. enjoys a contract from  
the Collaborative Mathematics
Project of La Caixa Foundation.
Research projects in which this work is included are
FIS2012-31324, from Spanish MINECO, 
and  2014SGR-1307, from AGAUR.

% Bibliografia
\addcontentsline{toc}{chapter}{Bibliography}
\bibliography{biblio}
\bibliographystyle{unsrt}

\end{document}